\documentclass[%
reprint,
 amsmath,amssymb,
 aps,
]{revtex4-1}

\usepackage[english]{babel}

\usepackage{letltxmacro}
\LetLtxMacro{\ORIGselectlanguage}{\selectlanguage}
\makeatletter
\DeclareRobustCommand{\selectlanguage}[1]{%
  \@ifundefined{alias@\string#1}
    {\ORIGselectlanguage{#1}}
    {\begingroup\edef\x{\endgroup
       \noexpand\ORIGselectlanguage{\@nameuse{alias@#1}}}\x}%
}
\newcommand{\definelanguagealias}[2]{%
  \@namedef{alias@#1}{#2}%
}
\makeatother
\definelanguagealias{en}{english}

\usepackage{graphicx}
\usepackage{dcolumn} 
\usepackage{bm} 
\usepackage{placeins} 

\def\Re{{\rm Re}}
\def\Im{{\rm Im}}

\def\p{\partial}
\def\rr{\mathbf{r}}

\begin{document}


\title{Nonadiabatic time-dependent density-functional theory at the cost of adiabatic local density approximation}
	
\author{D. R. Gulevich}
\affiliation{ITMO University, St. Petersburg 197101, Russia}

\author{Ya. V. Zhumagulov}
\affiliation{ITMO University, St. Petersburg 197101, Russia}

\author{A. V. Vagov}
\affiliation{ITMO University, St. Petersburg 197101, Russia}
\affiliation{Theoretische Physik III, Universit\"{a}t Bayreuth, 95440 Bayreuth, Germany}

\author{V. Perebeinos}
\affiliation{Department of Electrical Engineering, University at Buffalo, The State University of New York, Buffalo, NY 14260, USA}
\affiliation{ITMO University, St. Petersburg 197101, Russia}

\date{\today}

\begin{abstract}
We propose a computationally efficient approach to the
nonadiabatic time-dependent density functional theory (TDDFT) which is based on a representation of the frequency-dependent exchange correlation kernel as a response of a set of damped oscillators. The requirements to computational resources needed to implement our approach do not differ from those of the standard real-time TDDFT in the adiabatic local density approximation (ALDA). Thus, our result offers an exciting opportunity to take into account temporal nonlocality and memory effects in calculations with TDDFT in quantum chemistry and solid state physics for unprecedentedly low costs.
\end{abstract}

{\let\newpage\relax\maketitle}

TDDFT~\cite{Runge-Gross} has recently become a standard tool for studying electronic excitations in molecules, atomic clusters and solid state (see reviews~\cite{Botti-2007,Casida-2012,Ullrich-2014a,Maitra-2016} and books~\cite{Marques-2012,Ullrich-book}).
Its real-time formulation based on solving the time-dependent Kohn-Sham (TDKS) equation allows not only to obtain the linear excitation spectra, but also to study nonlinear dynamics under arbitrary time-dependent perturbations~\cite{Takimoto-2007,Lopata-2011,Yabana-2012,Castro-2012,Meng-2012,Schultze-2014,Bao-2015,Krieger-2015,Provorse-2016,Tancogne-Dejean-2017a,Tancogne-Dejean-2017,Pemmaraju-2018,Tancogne-Dejean-2018,Yost-2019,Yost-2019a}.

Implementations of real-time TDDFT available in the standard density-functional theory (DFT) packages~\cite{Soler-2002,qbox-rt,gpaw-rt,Andrade-2015,SALMON-2019} utilize the adiabatic approximation, which neglects the frequency-dependence of the exchange-correlation (xc) kernels introduced to capture exchange interactions and correlation effects of the electron subsystem.
Significant progress in development of the {\it nonadiabatic} TDDFT beyond the ALDA has been made by appreciating that the current $\mathbf{j}(\rr,t)$ rather than the density $n(\rr,t)$ could be used as the main variable in the xc functional of a nonadiabatic theory~\cite{Vignale-1996}. The idea has been cast in the form of hydrodynamic equations~\cite{Vignale-1996,VUC-PRL-1997,Ullrich-2002,Dobson-1997} referred to as the time-dependent current-density functional theory (TDCDFT) and a theory of deformations in the comoving Lagrangian frame~\cite{Tokatly-2005,Tokatly-2005a,Tokatly-2007}.
Demand for the nonadiabatic treatment comes from a number of research areas in quantum chemistry and solid state physics,
such as methods of optimal quantum control~\cite{qoc-book,qoc-1,qoc-2}, studies of multiple and double excitations in molecules~\cite{Maitra-2004,Zhang-2004,Gritsenko-2009,Romaniello-2009,Sangalli-2011,Sakkinen-2012} and physics of excitons in 2D materials~\cite{Novoselov-2016,2D-pero,Rivera-2016,Zhang-2015,Chernikov-2015,Turkowski-2009}.

The major obstacle limiting applications of the real-time TDCDFT is that the TDKS equation is essentially nonlocal and at each instant depends on values of the velocity field $\mathbf{u}(\rr,t)=\mathbf{j}(\rr,t)/n(\rr,t)$ in the entire past evolution. The need to store and process all the previous history makes numerical implementation of real-time TDCDFT impractical for realistic systems.

In this work, we show how to bypass this computational difficulty by presenting an efficient approach with the demand for computational resources similar to that for the real-time TDDFT-ALDA. Our method opens an exciting opportunity to take into account nonadiabatic effects in TDDFT calculations in quantum chemistry and solid state physics with only a marginal rise of the computing cost.
The proposed approach is based on the following representation of frequency-dependent xc kernels:
\begin{multline}
f_{xc}(n,\omega) = f_{xc}(n,\infty) \\ +
\frac{1}{2}\sum_{m=1}^M \bigg[
\frac{C_m(n)\,p_m(n)}{\omega-p_m(n)}
-\, \frac{C^{*}_m(n)\,p^{*}_m(n)}{\omega+p^{*}_m(n)}
\bigg],
\label{fxc-exp}
\end{multline}
where poles $p_m(n)$ and weights $C_m(n)$ are complex functions of electron density $n$.
Imaginary part of the poles satisfy the condition $\Im\, p_m(n) < 0$, which follows from causality of the response,
whereas real part of the weights are subjected to the sum rule
\begin{equation}
 \sum_{m=1}^M \Re\, C_m(n) = f_{xc}(n,\infty)-f_{xc}(n,0).
\label{constraint}
\end{equation}
Although the kernel in Eq. (\ref{fxc-exp}) may comprise infinitely many terms, in practice, there is always a finite number of dominant contributions which fully define the response. Furthermore, because the nonadiabatic kernels are known only approximately, a sufficiently accurate model can be constructed with a few dominant terms in Eq.~\eqref{fxc-exp}. As we  demonstrate below, even a kernel comprising a single term in the sum~\eqref{fxc-exp} is capable of reproducing  dynamics of interacting electron systems on a par with the significantly more complex models~\cite{GK-PRL-1985,GK-PRL-Errata,Iwamoto-PRB-1987,Bohm-1996,Conti-Nifosi-Tosi-1997,Nifosi-1997,Nifosi-1998, QV-PRB-2002}.

The form of Eq.~\eqref{fxc-exp} allows to draw an analogy with the classical models of optical susceptibility~\cite{Haug-2004}. Indeed, the representation Eq.~\eqref{fxc-exp} can be seen as a response of a set of damped Lorentz oscillators with the density-dependent natural frequencies $\omega_m(n)=|p_m(n)|$, damping parameters $-2\,\Im\, p_m(n)$ and oscillator strengths $\Re\,C_m(n)$. Thus, one can interpret Eq.~\eqref{fxc-exp} as an effective interaction of the electron subsystem with a set of independent oscillators responsible for the temporal nonlocality of the TDKS equation. 
We refer to this model as the {\it oscillator model for xc kernels} (OMXC).

\begin{figure}[t!]
	\begin{center}
\includegraphics[width=3.2in]{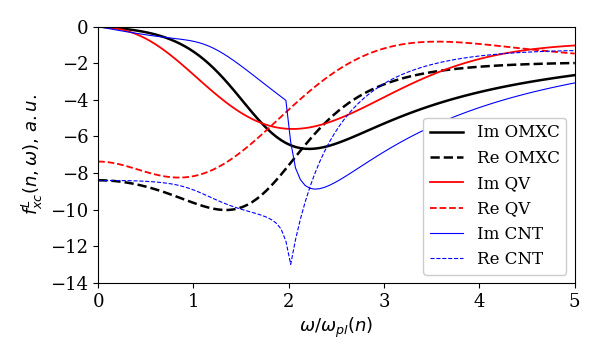}
		\caption{\label{kernel}
The single-oscillator OMXC kernel (thick black lines) given by the Eq.~\eqref{p1}. Parametrizations of Qian-Vignale~\cite{QV-PRB-2002} (QV, red lines) and
Conti-Nifosi-Tosi~\cite{Conti-Nifosi-Tosi-1997, Nifosi-1998} (CNT, blue lines) for 3D electron gas are also shown. All kernels are evaluated at the electron density $n$ which corresponds to the Wigner-Seitz radius $r_s=3$ Bohr.
		}		
	\end{center}		
\end{figure}

The OMXC has the following key properties which make it particularly useful for practical implementation of the nonadiabatic TDDFT:

(i) Both real and imaginary parts of the OMXC kernels are defined by explicit complex function of the complex frequency.
In contrast, the standard parametrizations~\cite{GK-PRL-1985,GK-PRL-Errata,Iwamoto-PRB-1987,Bohm-1996,Conti-Nifosi-Tosi-1997,Nifosi-1997,Nifosi-1998, QV-PRB-2002} of the imaginary part of xc kernels require a careful analytic continuation into the lower complex half-plane~\cite{Ullrich-Vignale-PRB-1998} and 
evaluation of complicated integrals in order to find the real part~\cite{dfc-2019}.

(ii) Imaginary and real parts of the quantity $f_{xc}(n,\omega)-f_{xc}(n,\infty)$ given by the Eq.~\eqref{fxc-exp}
satisfy the Kramers-Kronig relations.

(iii) When transformed to the time-domain, Eq.~\eqref{fxc-exp} yields an explicit function of time which satisfies the causality.

(iv) A simple analytic structure of the Eq.~\eqref{fxc-exp} admits an intuitive interpretation providing an insight into the underlying physics. In some cases, positions of the poles $p_m(n)$ can be guessed from the relevant physical processes and, vice versa, location of the poles may explain the physics behind the nonadiabatic xc kernels.

(v) Most importantly, the OMXC
enables one to construct highly efficient numerical schemes for the real-time TDCDFT. In what follows, we will
elaborate the foundations for the efficient implementations of real-time TDCDFT. We will focus on 3D electron systems, although our approach can be applies to 2D systems equally well.

The bottleneck of the real-time TDCDFT
is evaluation of
the xc viscoelastic stress tensor~\cite{VUC-PRL-1997}:
\begin{multline}
\sigma^{xc}_{ij}(\mathbf{r},t)=\int_0^t dt' \bigg\{ \eta_{xc}(\bar{n},t-t') \Big[
 \p_i u_j(\rr,t')  + \p_j u_i(\rr,t') \\
 - \frac{2}{3}\nabla\cdot \mathbf{u}(\rr,t')\delta_{ij}
\Big]
+ \zeta_{xc}(\bar{n},t-t') \nabla\cdot\mathbf{u}(\rr,t')\delta_{ij}
\bigg\},
\label{sxc}
\end{multline}
where
$\mathbf{u}(\mathbf{r},t)$
is the velocity field and we assume that at $t<0$ the electron system is in the ground state. The average density $\bar{n}$ which enters Eq.~\eqref{sxc} can be evaluated at either $t$ or $t^\prime$ as the difference between $n(\rr,t)$ and $n(\rr,t')$ can be neglected within the Vignale-Kohn approximation~\cite{Vignale-1996,WU-PRL-2005,Ullrich-Tokatly-PRB-2006}.
The time-dependent kernels $\eta_{xc}(n,t)$ and $\zeta_{xc}(n,t)$ are the Fourier transforms of the complex viscosity coefficients
\begin{equation}
\begin{split}
\eta_{xc}(n,\omega)&=\frac{in^2}{\omega+i0}\;f_{xc}^T(n,\omega),
\\
\zeta_{xc}(n,\omega)&=\frac{in^2}{\omega+i0}\left[ f_{xc}^L(n,\omega)-\frac43 f_{xc}^T(n,\omega)-\frac{d^2\epsilon_{xc}(n)}{dn^2} \right],
\end{split}
\label{eta-zeta}
\end{equation}
where the label $L$ ($T$) stands for the longitudinal (transverse)  component of the kernel, and $\epsilon_{xc}(n)$ is the  xc energy density
of the homogeneous electron gas.
The standard approach~\cite{WU-PRL-2005,Ullrich-Tokatly-PRB-2006,Ullrich-2006a} of solving the TDKS equation in the real-time TDCDFT is to calculate the time integral for the stress tensor in Eq.~\eqref{sxc} directly using the time-dependent kernels $\eta_{xc}(n,t)$ and $\zeta_{xc}(n,t)$.
However, this brute-force approach puts an enormous load on the computational resources. Indeed, to propagate the TDKS equation, one needs to store
the previous evolution of the velocity field, while using it in evaluations of the integrals~\eqref{sxc} at every time step. To keep the computation feasible, one is forced to introduce a cutoff for the memory depth to prevent an unbound rise of the computer memory and computing time~\cite{Kurz-2006,Ullrich-2006a}. Often, one resorts to the instantaneous (Markovian) approximation~\cite{Escartin-2015,Dinh-2018} by neglecting the very dependence on the evolution history.

With the help of the OMXC one can significantly reduce both the computing time and the extensive load on the computer memory without any compromise for the treatment of nonadiabatic effects. In contrast to the instantaneous approximation, our method enables one to take into account arbitrary long memory of the evolution. The main idea is to avoid evaluation of the costly time integrals in Eq.\eqref{sxc}
and of the inverse Fourier transforms of Eq.~\eqref{eta-zeta}
by introducing auxiliary dynamical equations associated with individual oscillators of the OMXC.
Substituting Eq.~\eqref{fxc-exp} for both longitudinal and transverse components of the xc kernel in Eq.~\eqref{eta-zeta}, yields:
\begin{multline}
\eta_{xc}(n,\omega)= \frac{in^2}{2}\sum_{m=0}^M
\left[ \frac{C^{T}_m(n)}{\omega-p^{T}_m(n)} + \frac{C^{T*}_m(n)}{\omega+p^{T*}_m(n)} \right],
\label{eta-sum}
\end{multline}
and
\begin{multline}
\zeta_{xc}(n,\omega) = -\,\frac43 \eta_{xc}(n,\omega) \\
+ \frac{in^2}{2}\sum_{m=0}^M \left[ \frac{C^{L}_m(n)}{\omega-p^{L}_m(n)} + \frac{C^{L*}_m(n)}{\omega+p^{L*}_m(n)} \right],
\label{zeta-sum}
\end{multline}
where we introduced the terms at $m=0$, such that $p_0^{L,T}(n) = -i0$ and
\begin{equation}
C^L_0(n) = f_{xc}^L(n,0)-\frac{d^2\epsilon_{xc}(n)}{dn^2}, \quad C^{T}_0(n) = f_{xc}^T(n,0).
\end{equation}
Transforming~\eqref{eta-sum} and \eqref{zeta-sum} to the time-domain and substituting to Eq.~\eqref{sxc},
we reduce the expression for  xc stress tensor to
\begin{multline}
\sigma^{xc}_{ij}(\mathbf{r},t)
= n^2 \sum_{r=L,T} \sum_{m=0}^M \Re \int_0^t C_m^r(\bar{n})
\\
\times e^{-ip_m^r(\bar{n})(t-t')} \mu^{r}_{ij}(\rr,t')\, dt',
\label{sxc-int}
\end{multline}
where we introduced
\begin{equation}
\begin{split}
\mu^L_{ij}(\rr,t) &= \nabla\cdot\mathbf{u}(\rr,t)\delta_{ij},
\\
\mu^T_{ij}(\rr,t) &=  \p_i u_j(\rr,t) + \p_j u_i(\rr,t) - 2\nabla\cdot \mathbf{u}(\rr,t)\delta_{ij}.
\end{split}
\end{equation}
Employing the ambiguity of the Vignale-Kohn approximation~\cite{Vignale-1996}, according to which $\bar{n}$ in Eq.~\eqref{sxc-int} is either $n(\rr,t)$ or $n(\rr,t')$, we can rewrite the expression for the xc stress tensor in the form
\begin{equation}
\sigma^{xc}_{ij}(\mathbf{r},t)
= n^2 \sum_{r=L,T} \sum_{m=0}^M \Re \left[ C_m^r(n(\rr,t)) \mathcal{M}^r_{mij}(\rr,t) \right],
\label{sxc-new}
\end{equation}
where $\mathcal{M}^r_{mij}(\rr,t)$ are memory variables which satisfy the evolution equation
\begin{equation}
\frac{\p}{\p t}\mathcal{M}^r_{mij}(\rr,t) = \mu^{r}_{ij}(\rr,t) - i p_m^r(n(\rr,t))\,\mathcal{M}^r_{mij}(\rr,t),
\label{ode}
\end{equation}
with the initial condition $\mathcal{M}^r_{mij}(\rr,0)=0$.

\begin{figure}[t!]
	\begin{center}
\includegraphics[width=3.3in]{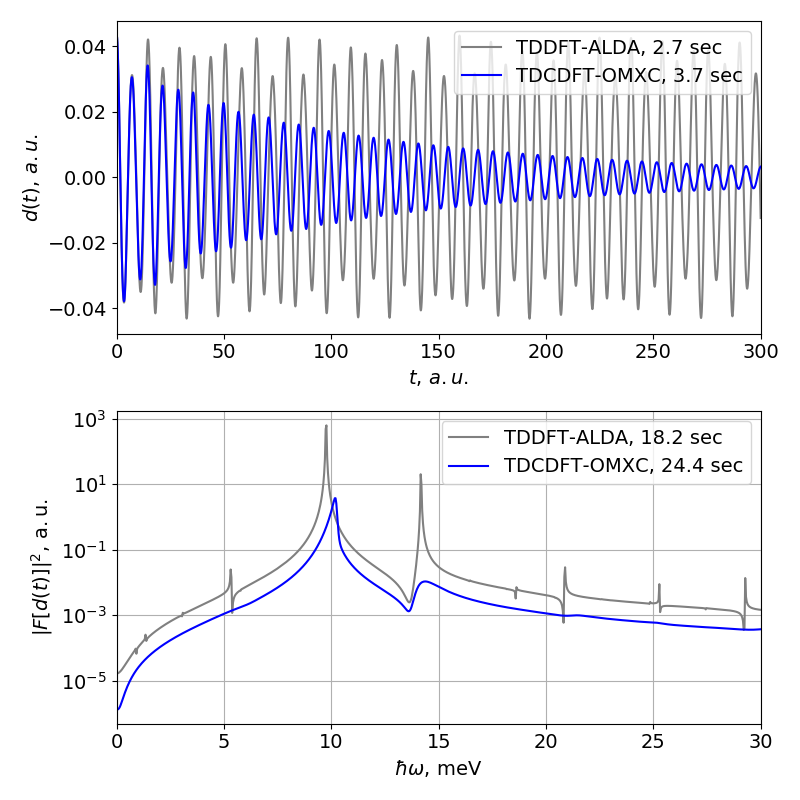}
		\caption{\label{fig2}
		Performance of the TDCDFT-OMXC (blue line) applied to calculate the nonlinear dynamics of dipole moment in $\rm GaAs/Al_{0.3}Ga_{0.7}As$ quantum well, in comparison to the the standard TDDFT-ALDA (grey line). The spectrum (lower panel) is calculated for the time evolution of the dipole moment during 2000 a.u.
		The indicated mean computation time is measured in a serial run on a single core of Intel~i7 processor.
		}		
	\end{center}		
\end{figure}

The memory variables $\mathcal{M}^{r}_{mij}(\rr,t)$ are associated with individual OMXC oscillators and carry information about previous evolution of the system, holding, in principle, an infinite memory of $\mu^{r}_{ij}(\rr,t)$. Eqs.~\eqref{sxc-new},~\eqref{ode}, which have to be solved in addition to the TDKS equation, eliminate the need of both storing the past evolution and evaluating the integrals in Eq.~\eqref{sxc}.
Thus, this approach lays the foundations for highly efficient numerical schemes, where all memory effects are taken into account by propagating auxiliary time-local differential equations (cf. numerical schemes used in the Josephson physics~\cite{OSZ,Gulevich-MTT}).



It is evident that given $p^r_m(n)$ and $C^r_m(n)$,
the computational overhead of finding $\sigma^{xc}_{ij}(\rr,t)$ according to the Eqs.~\eqref{sxc-new},~\eqref{ode} is minor compared to solving the TDKS equation itself. Therefore, the only factor which limits the computational efficiency of this approach is evaluations of  $p^r_m(n)$ and $C^r_m(n)$ from the electron density $n(\rr,t)$. However, such computational costs are comparable to evaluation of the ALDA  exchange-correlation potential. Moreover, the numerical effort can be further reduced, if the expressions for $p^r_m(n)$ and $C^r_m(n)$ involve powers of the electron density, such as $n^{1/3}$ or similar. In this case, their values can be reused from evaluation of the adiabatic component of the xc vector potential. Thus, the performance of the real-time TDCDFT based on the proposed approach (TDCDFT-OMXC) promises to be comparable to that of the standard real-time TDDFT-ALDA.
As a proof of concept, we will construct a simple OMXC kernel of the form~\eqref{fxc-exp} and apply it to the benchmark problem of collective intersubband excitations in a quantum well~\cite{Ullrich-Vignale-PRB-1998,WU-PRL-2005}.

To derive the single-oscillator OMXC kernel, we proceed as follows.
We will focus on the longitudinal component $f_{xc}^L(n,\omega)$ which we will need in our calculation of dynamics of intersubband excitations, although, the same arguments apply to the transverse component $f_{xc}^T(n,\omega)$ as well. Using the prediction that $\Im\,f_{xc}^L(n,\omega)$ has a peak at twice the plasmon frequency $\omega_{pl}(n)$ ~\cite{Bohm-1996,Conti-Nifosi-Tosi-1997}, we set $|p^L_1(n)|=2\,\omega_{pl}(n)$.
Despite the theoretical efforts~\cite{Conti-Nifosi-Tosi-1997,Nifosi-1998,QV-PRB-2002},
the peak width $-2\,\Im\, p^L_m(n)$ at the double plasmon frequency is much less known.
The perturbation theory~\cite{STLS,Conti-Nifosi-Tosi-1997,Nifosi-1998} predicts the low-frequency side of the double plasmon peak to scale with the plasma frequency $\omega_{pl}(n)$. Thus, for the single-pole OMXC kernel we take:
\begin{equation}
|p^L_1(n)| = 2\,\omega_{pl}(n), \quad \Im\,p^L_1(n) = -\,\gamma_1^L \omega_{pl}(n),
\label{p1}
\end{equation}
with the proportionality parameter $\gamma_1^L$.
Given Eq.~\eqref{p1}, $C_1^L(n)$ is uniquely determined by the sum rule in Eq.~\eqref{constraint} and the infra-red asymptotics of the kernel,
\begin{equation}
-\, \Im\, \frac{C_1^L(n)}{p_1^L(n)} =
\lim_{\omega\to 0}
\frac{\Im\,f_{xc}^L(n,\omega)}{\omega}.
\label{derivative}
\end{equation}
The right hand side of~\eqref{derivative} can be evaluated from the perturbation theory~\cite{QV-PRB-2002} which
predicts a small but nonzero tangent of $\Im\,f_{xc}^L(n,\omega)$ at $\omega=0$.
The frequency dependence of the OMXC kernel~\eqref{p1}
is shown in Fig.~\ref{kernel} alongside the standard parametrizations of
Conti-Nifosi-Tosi~\cite{Conti-Nifosi-Tosi-1997, Nifosi-1998} (CNT)
and Qian-Vignale~\cite{QV-PRB-2002} (QV).
Note, that the single-oscillator OMXC kernel in Fig.~\ref{kernel} at $\gamma_1^L=1$ is intermediate between the CNT and QV parametrizations. Despite its simplicity, our single-oscillator OMXC kernel correctly describes the low-frequency asymptotics and the qualitative features at the double plasmon frequency predicted earlier~\cite{Conti-Nifosi-Tosi-1997, Nifosi-1998}.

\begin{figure}[t!]
	\begin{center}
\includegraphics[width=3.3in]{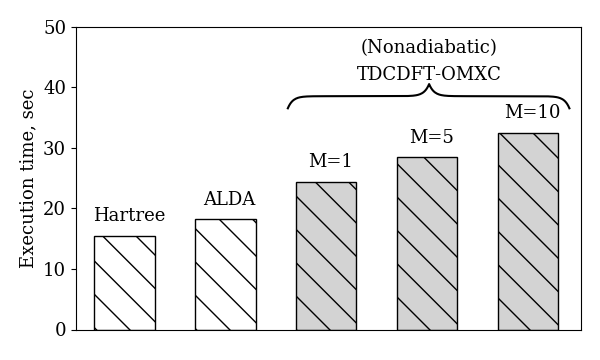}
		\caption{\label{timing}
The mean computation time required to propagate the electron density
in $\rm GaAs/Al_{0.3}Ga_{0.7}As$ quantum well
from $t=0$ to $2000$ a.u. with time step $0.02$ a.u.
The first two columns correspond to the Hartree and ALDA calculations, whereas the last three is our TDCDFT-OMXC implementation using kernels~\eqref{fxc-exp} with different $M$. The overhead due to the account of nonadiabatic effects using the single-oscillator kernel is about 35\% of the time required for the ALDA calculation. The benchmarks were obtained for serial execution on a single core of Intel~i7 processor.
		}		
	\end{center}		
\end{figure}

We verify the performance of our TDCDFT-OMXC approach with the kernel~\eqref{p1} on a benchmark problem of intersubband excitations in $\rm GaAs/Al_{0.3}Ga_{0.7}As$ quantum well~\cite{Ullrich-Vignale-PRB-1998,WU-PRL-2005}.
Our calculations in the linear regime are consistent with the linear response theory of Refs.~\cite{Ullrich-Vignale-PRB-1998}: the value for the mode frequency $10.23\pm 0.02$ meV obtained using the TDCDFT-OMXC in the linear regime agrees well with the results in Ref.~\cite{Ullrich-Vignale-PRB-1998}.
In our calculations of the nonlinear dynamics we assumed that the initial state of the system is its ground state in the electric field 0.5 mV/nm. At $t=0$ the electric field is switched off and the electron liquid in the quantum well evolves freely in time, as seen in the
evolution of the dipole moment $d(t)$ presented in the Fig.~\ref{fig2}.
The arising spectrum of frequencies associated with the nonlinear evolution is shown in the lower panel of Fig.~\ref{fig2}. The most remarkable is the computational costs
at which our nonadiabatic results were obtained:
the computing time using the TDCDFT-OMXC is only about 35\% larger than the time required to propagate the electron density using the standard TDDFT-ALDA~\footnote{To obtain the results presented in Fig.~\ref{fig2} we used the single-oscillator kernel given by Eqs.~\eqref{p1} assuming zero tangent of $\Im f^L_{xc}(n,\omega)$ at $\omega=0$ in Eq.~\eqref{derivative}. The use of the exact tangent~\eqref{derivative} from the perturbation theory~\cite{QV-PRB-2002} takes about 30\% longer execution time, but has no observable effects on the dynamics of the quantum well as compared to the simpler version of the kernel}.
In contrast, the brute-force approach with the direct evaluation of the memory integrals in Eq.~\eqref{sxc} takes an incomparably longer time, exceeding the timing of ALDA calculation by few orders of magnitude.
Further studies of intersubband excitations are provided in the Supplemental Material
~\footnote{See Supplemental Material at [URL will be inserted by publisher] for further details on the derivation of OMXC kernels and benchmarks of the TDCDFT-OMXC approach.}.

In our derivation of the single-oscillator kernel~\eqref{p1} we ignored the high-frequency asymptotics, which is a valid assumption for the low-frequency dynamics, $\omega<E_F$, where $E_F$ is the Fermi energy.
It is also possible to construct OMXC kernels valid in the entire frequency range.
Indeed, while formally the OMXC can never satisfy the exact high-frequency asymptotics $\Im f_{xc}(n,\omega) \sim \omega^{-3/2}$~\cite{Holas-Singwi}, in practice, the frequency range of interest is always limited from above by a finite value.
Therefore, it is possible to construct an approximation to the exact high-frequency xc kernel using a finite number of oscillators in the OMXC. In the Supplemental Material~\cite{Note2} we provide an example of a three-oscillator OMXC kernel
which satisfies the high-frequency asymptotics up to frequencies 100 $E_F$ with the relative error below $1\%$.

Because constructing more precise xc kernels will likely involve $M>1$ contributing terms in Eq.~\eqref{fxc-exp}, we analyze how our TDCDFT-OMXC approach scales with the number of oscillators.
The computing time required to simulate the nonlinear dynamics of electron liquid in $\rm GaAs/Al_{0.3}Ga_{0.7}As$ quantum well using kernels with different values of $M$ is shown in Fig.~\ref{timing} alongside performances of the bare Hartree and TDDFT-ALDA. Note that the execution time grows relatively slow with $M$, so that even the sophisticated 10-oscillator OMXC kernel can be used in the nonadiabatic calculations at the computational cost of only twice the timing of the standard TDDFT-ALDA.

To conclude, we propose a computationally efficient approach to the nonadiabatic TDCDFT which enables one to replace the costly memory integrals 
over the whole previous time evolution with auxiliary differential equations, thereby diminishing the computational costs to that of the standard TDDFT-ALDA calculation. Besides, the OMXC kernels used in our approach have a number of useful properties: they are defined in the whole complex frequency plane, satisfy the causality, give an explicit expression for the real and imaginary parts, and have an intuitively transparent structure familiar from the standard theory of optical response.
We expect that our TDCDFT-OMXC approach
will open exciting opportunities for solving computationally prohibitive tasks in quantum chemistry and solid state physics beyond the ALDA.

{\it Acknowledgments.}
The work is supported by the Russian Science Foundation under the grant 18-12-00429. The authors would like to thank Ilya Tokatly for helpful discussions.

\bibliography{bibliography}

\pagebreak
\widetext
\begin{center}
\textbf{\large Supplemental Material for ``Nonadiabatic time-dependent density-functional theory at the cost of adiabatic local density approximation"}
\end{center}
\setcounter{equation}{0}
\setcounter{figure}{0}
\setcounter{table}{0}
\setcounter{page}{1}
\makeatletter
\renewcommand{\theequation}{S\arabic{equation}}
\renewcommand{\thefigure}{S\arabic{figure}}
\renewcommand{\bibnumfmt}[1]{[S#1]}
\renewcommand{\citenumfont}[1]{S#1}

\section*{Single-oscillator kernel}

For the strength of the single-oscillator OMXC kernel given by Eq.(12) of the main text we obtain from the sum rule (2) and the constraint (13):
\begin{equation}
C_1(n) = \frac{p_1(n)}{\Re\,p_1(n)}
\Big[f_{xc}(n,\infty)-f_{xc}(n,0) - i\, p_1^{*}(n)\, D(n) \Big],
\end{equation}
where 
\begin{equation}
D(n)= \lim_{\omega\to 0}\frac{\Im\,f_{xc}(n,\omega)}{\omega}
\end{equation}
is given by Eq.(15) of Ref.~\cite{supp-QV-PRB-2002}.

In our numerical calculations for the dynamics of the dipole moment in a quantum well  presented in Fig.2 of the main paper we have set $D(n)=0$.  The use of the exact tangent of $\Im\,f_{xc}(n,\omega)$ at $\omega=0$
calculated by~\cite{supp-QV-PRB-2002} takes about 30\% longer execution time (that is, 70\% above ALDA), but has no observable effects on the dynamics of the quantum well as compared to the simpler version
of the kernel used in calculation in the main text.

\section*{High-frequency kernels}

The second order perturbation theory~\cite{supp-Glick-Long,supp-GK-PRL-1985,supp-Holas-Singwi} predicts the high-frequency $\omega\gg E_F$ asymptotic expansion for the xc kernel of the 3D electron gas to be:
\begin{equation}
\Im\, f_{xc}(n,\omega) \approx -\,\frac{23\pi}{15}\,\omega^{-3/2}.
\label{w32-theory}
\end{equation}
In the following we will assume the high-frequency asymptotics is given by the following general form:
\begin{equation}
\Im\, f_{xc}(n,\omega) \approx c(n)\,\left[\frac{\omega}{E_F(n)} \right]^{-3/2}.
\label{w32}
\end{equation}
Expanding the imaginary part of Eq.(1) at $\omega\to\infty$ we obtain:
\begin{equation}
\Im\, f_{xc}(n,\omega) = 
\frac{E_F(n)}{\omega} \sum_{m=1}^M \frac{\Im\left[C_m(n)\,p_m(n)\right]}{E_F(n)} +
\left[\frac{E_F(n)}{\omega}\right]^3\sum_{m=1}^M  
\frac{\Im\left[ C_m(n)\,p^{3}_m(n)\right]}{E_F^3(n)} + ....
\label{fxc-asympt}
\end{equation}
It is, therefore, convenient to set $\Im\left[C_m(n)\,p_m(n)\right]=0$ for all oscillators $m=1...M$, so that the terms of the order ${\cal O}(1/\omega)$ are absent in the asymptotic expansion~\eqref{fxc-asympt}. 

Obviously, within the OMXC, the asymptotics~\eqref{w32} cannot be exactly satisfied in the limit $\omega\to\infty$ as it would require and infinite number of oscillators. Nevertheless, Eq.~\eqref{w32} can be satisfied inside 
a finite interval $1\ll \tilde\omega_{min} <\omega/E_F(n)<\tilde\omega_{max}$
up to a given precision using a finite number of terms. Consider a set of overdamped oscillators with the poles at
\begin{equation}
p_m(n)=r_m(n) -i\gamma_m E_F(n),\quad \gamma_m\gg 1,
\label{pms}
\end{equation}
which we will use to model the asymptotics~\eqref{w32}. Here, the precise value of $\Re\,p_m(n) = r_m(n)\sim E_F(n)$ is irrelevant due the condition $\gamma_m\gg 1$. The strengths $C_m(n)$ for the high-frequency oscillators are then defined from the constraints 
\begin{equation}
\Im\left[C_m(n)\,p_m(n)\right]=0, 
\quad 
\frac{\Im\left[C_m(n)\,p_m(n^3)\right]}{E_F^3(n)}=c_m\,c(n),
\label{Cmeqs}
\end{equation}
which yield
\begin{equation}
C_m(n) = \frac{c_m\,c(n)\,E_F^3(n)}{p_m(n)\, \Im\,p_m^2(n)},
\label{Cms}
\end{equation}
where values of real density-independent coefficients $c_m$ entering Eqs.~\eqref{Cmeqs},~\eqref{Cms} are such that the resulting OMXC kernel satisfies the asymptotics~\eqref{w32} to the desired precision. 
Using~\eqref{Cms} and comparing to~\eqref{w32} 
we obtain the cost function for relative errors, which has to be minimized with respect to parameters $c_m$ and $\gamma_m$,
\begin{equation}
R(c_1,...,c_M,\gamma_1,...,\gamma_M) = \int_{\tilde\omega_{min}}^{\tilde\omega_{max}} 
d\tilde\omega\,  \left( \frac12 \sum_{m=1}^M \frac{c_m}{\gamma_m} \;
\Im\left[ \frac{\tilde\omega^{3/2}}{\gamma_m^2 - \tilde\omega^2 - 2i\gamma_m\tilde\omega}\right]
\, - \, 1
\right)^2,
\label{R}
\end{equation}
where $\tilde\omega=\omega/E_F(n)$ and the real parts of the poles~\eqref{pms} are neglected due to $\gamma\gg 1$. Note, that the cost function~\eqref{R} is independent of the density. Therefore, the coefficients $c_m$ are universal and their calculation can be done only once for the given interval and the number of oscillators.

\begin{figure}[h!]
	\begin{center}
\includegraphics[width=4in]{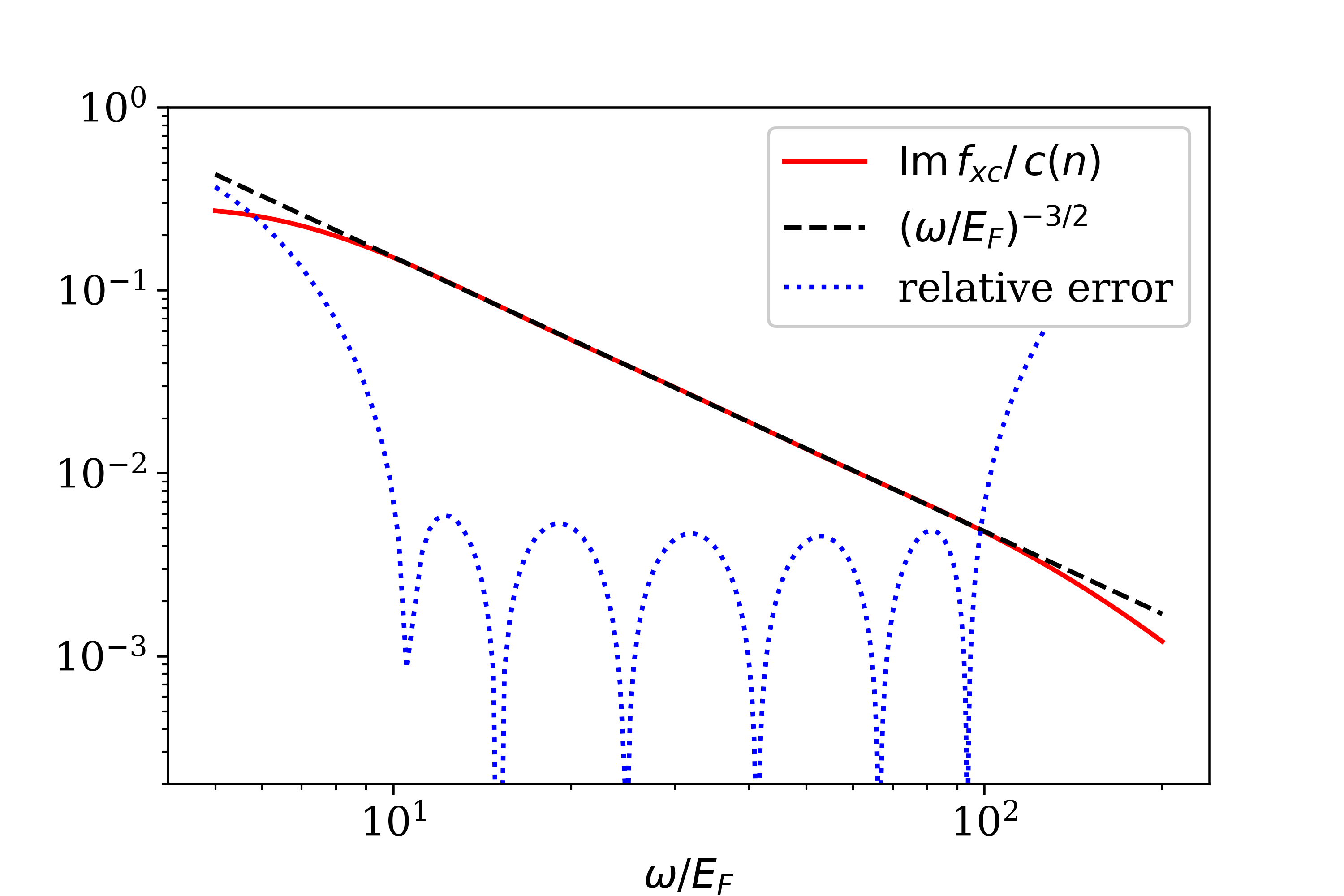}
		\caption{\label{fit} 
Three-oscillator OMXC kernel (red solid line) optimized for the range 
$10<\tilde \omega < 100$ and the exact high-frequency asymptotics~\eqref{w32-theory} (dashed line). 
 The density-dependent function $c(n)$ is defined in Eqs.~\eqref{w32-theory},~\eqref{w32} and given by $c(n)=-(23\pi/15)\,E_F^{-3/2}$. 
The dotted blue line shows the relative error between the three-oscillator fit and the asymptotics~\eqref{w32-theory}.
		}		
	\end{center}		
\end{figure}

Given the values of $c_m$ and $\gamma_m$, obtained from the minimization of~\eqref{R}, the multiple-oscillator OMXC kernel which satisfies the high-frequency asymptotics~\eqref{w32} is defined by the poles~\eqref{pms} and the weights~\eqref{Cms}. In Fig.~\ref{fit} we provide example of a three-oscillator kernel obtained by minimization of~\eqref{R} in the range $10<\tilde \omega < 100$.
Parameters of the kernel are: $\gamma_1=7.064$, $\gamma_2=27.63$, $\gamma_3=97.00$, $c_1=60.474$, $c_2=313.28$, $c_3=2497.81$.

\section*{Intersubband transitions in quantum well}

Numerical solution of the TDKS equation for TDCDFT-OMXC and TDDFT-ALDA was implemented in the Kohn-Sham basis using the self-consistent Crank-Nicolson algorithm~\cite{supp-Ullrich-book}. The code for numerical calculations was implemented in C++.

\begin{figure}[h!]
	\begin{center}
\includegraphics[width=5.5in]{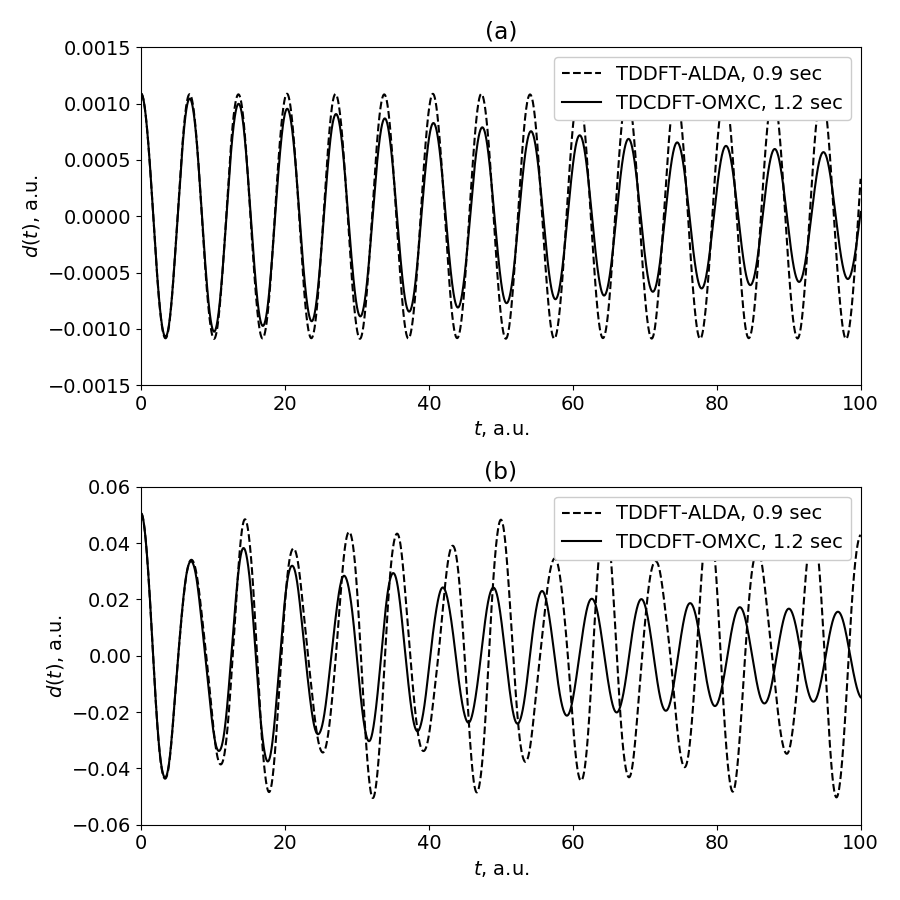}
		\caption{\label{figS1} 
Dynamics of a dipole moment in $\rm GaAs/Al_{0.3}Ga_{0.7}As$ quantum well~\cite{supp-WU-PRL-2005} calculated using our implementation of TDCDFT (solid line). The TDDFT-ALDA result is presented by the dashed line. Parameters of the quantum well and the system setup correspond to the  Ref.~\cite{supp-WU-PRL-2005}: the initial state is taken the ground state in the electric field 0.01 mV/nm (a) and 0.5 mV/nm (b). At $t=0$ the electric field is switched off and the free evolution of the dipole moment is calculated. The execution time required to perform the calculation is indicated on the legend.
		}		
	\end{center}		
\end{figure}

Apart from the results presented in the main paper, we have applied our implementation of TDDFT to calculate dynamics of the dipole moment in the quantum well with parameters taken from Ref.~\cite{supp-WU-PRL-2005}. Our results calculated using the single-oscillator kernel are presented in Fig.~\ref{figS1}. 

\FloatBarrier


\providecommand{\noopsort}[1]{}\providecommand{\singleletter}[1]{#1}%

\end{document}